%% file: main.tex
\documentclass[letter, 10 pt, conference]{ieeeconf}  

\IEEEoverridecommandlockouts 
\usepackage[english]{babel}
\usepackage[utf8]{inputenc}
\usepackage{times} 
\usepackage{cite}

\DeclareMathAlphabet{\pazocal}{OMS}{zplm}{m}{n}

\usepackage{hyperref}
\usepackage[nolist]{acronym}

\usepackage{multirow}

\usepackage{siunitx}
\usepackage{graphicx} 
\usepackage{epsfig} 
\usepackage{pgfplots}
\usepackage{caption}
\usepackage[font=footnotesize]{subcaption}

\usepackage{float}

\usepackage[export]{adjustbox}

\usepackage{xcolor}
\definecolor{myYellow}{rgb}{0.93,0.69,0.13}
\definecolor{myPurple}{rgb}{0.49,0.18,0.56}
\definecolor{myGreen}{RGB}{0 204 0}

\usepackage{mathtools}
\usepackage{nicefrac}
\usepackage{amsmath}
\usepackage{amssymb}
\usepackage{bm} 

\usepackage{etoolbox}
\makeatletter%
\AfterPreamble{%
	\usepackage{hyperref}%
	\let\oldhypertarget\hypertarget%
	\renewcommand{\hypertarget}[2]{%
		\oldhypertarget{#1}{#2}%
		\protected@write\@mainaux{}{%
			\string\expandafter\string\gdef%
			\string\csname\string\detokenize{#1}\string\endcsname{#2}%
		}%
	}%
	\newcommand{\myhyperlink}[1]{%
		\hyperlink{#1}{\csname #1\endcsname}%
	}%
}
\makeatother%

\newcounter{Remark}
\newcounter{Definition}
\newcounter{Problem}
\usepackage{algorithm}
\usepackage[noend]{algpseudocode}

\makeatletter
\def\BState{\State\hskip-\ALG@thistlm}
\makeatother

\usepackage{tikz}
\pgfdeclarelayer{foreground}
\pgfsetlayers{background, main, foreground}
\usetikzlibrary{quotes, angles, backgrounds, arrows, automata, shapes, positioning, calc, through, spy, decorations.pathreplacing, decorations.markings, arrows.meta, automata, petri}

\tikzset{
    imglabel/.style={
      rectangle,
      inner sep=2pt,
      text=black,
      minimum height=1em,
      text centered,
      fill=white,
      fill opacity=1.0,
      text opacity=1,
      anchor=south west,
    },
  }
\tikzset{
	state/.style={
		rectangle,
		draw=black, very thick,
		minimum height=1.0em,
		text centered,
	},
}
\tikzset{
  on each segment/.style={
    decorate,
    decoration={
      show path construction,
      moveto code={},
      lineto code={
        \path [#1]
        (\tikzinputsegmentfirst) -- (\tikzinputsegmentlast);
      },
      curveto code={
        \path [#1] (\tikzinputsegmentfirst)
        .. controls
        (\tikzinputsegmentsupporta) and (\tikzinputsegmentsupportb)
        ..
        (\tikzinputsegmentlast);
      },
      closepath code={
        \path [#1]
        (\tikzinputsegmentfirst) -- (\tikzinputsegmentlast);
      },
    },
  },
  mid arrow/.style={postaction={decorate,decoration={
        markings,
        mark=at position .5 with {\arrow[#1]{stealth}}
      }}},
}

\graphicspath{{./figures/}}

\newcommand\copyrighttext{%
    \small \begin{center} \color{red} \textcopyright\,2024 IEEE. Accepted for presentation to the ``2024 Open Source Modelling and Simulation of Energy Systems" , Vienna, Austria. Personal use of this material is permitted. Permission from IEEE must be obtained for all other uses, in any current or future media, including reprinting/republishing this material for advertising or promotional purposes, creating new collective works, for resale or redistribution to servers or lists, or reuse of any copyrighted component of this work in other works. \end{center}}
\newcommand\copyrightnotice{%
	\begin{tikzpicture}[remember picture,overlay]
	\node[anchor=south,yshift=25.6cm] at (current page.south) 
	{\color{red}\fbox{\parbox{\dimexpr\textwidth-\fboxsep-\fboxrule\relax}{\copyrighttext}}};
	\end{tikzpicture}%
}

\title{\copyrightnotice \LARGE \bf Integrating Power-to-Heat Services in Geographically Distributed Multi-Energy Systems: A Case Study from the ERIGrid 2.0 Project}

\author{Giuseppe Silano$^{1,2}$, Evangelos Rikos$^{3}$, Vetrivel Rajkumar$^{4}$, Oliver Gehrke$^{5}$, Tesfaye Amare Zerihun$^{6}$, \\Carmine Rodio$^{1}$, and Riccardo Lazzari$^{1}$
    \thanks{$^1$G. Silano, R. Lazzari, and C. Rodio are with the Ricerca sul Sistema Energetico, Milan, Italy, (emails: {\tt \{name.surname\}@rse-web.it}).}
    \thanks{$^2$G. Silano is with the Czech Technical University in Prague, Czech Republic, (email: {\tt giuseppe.silano@fel.cvut.cz}).}
    \thanks{$^3$E. Rikos is with the Centre For Renewable Energy Sources and Saving, Athens, Greece, (email: {\tt vrikos@cres.gr}).}
    \thanks{$^4$V. Rajkumar is with the Delft University of Technology, Delft, The Netherlands, (email: {\tt v.subramaniamrajkumar@tudelft.nl}).}
    \thanks{$^5$O. Gehrke is with the Technical University of Denmark, Roskilde, Denmark, (email: {\tt olge@dtu.dk}).}
    \thanks{$^5$T. A. Zerihun is with the SINTEF Energy AS, Trondheim, Norway (email: {\tt tesfaye.zerihun@sintef.no}).}
    \thanks{This work received funding under the European Community's Horizon 2020 Program (H2020/2014-2020) in project ``ERIGrid 2.0'' (Grant Agreement No. 870620).}
}

\begin{document}

\maketitle
\thispagestyle{empty} 
\pagestyle{empty} 


\begin{acronym}
    \acro{BESS}[BESS]{Battery Energy Storage System}
    \acro{CHP}[CHP]{Combined Heat and Power}
    \acro{CRES}[CRES]{Center For Renewable Energy Sources}
    \acro{CSC}[CSC]{Centralized Supervisory Controller}
    \acro{DHN}[DHN]{District Heating Network}
    \acro{DSO}[DSO]{Distribution System Operator}
    \acro{DTU}[DTU]{Technical University of Denmark}
    \acro{EHP}[EHP]{Electrical Heat Pump}
    \acro{ESS}[ESS]{Energy Storage Systems}
    \acro{GDS}[GDS]{Geographically Distributed Simulations}
    \acro{HIL}[HIL]{Hardware-in-the-Loop}
    \acro{PCC}[PCC]{Point of Common Coupling}
    \acro{PV}[PV]{Photovoltaic}
    \acro{OS}[OS]{Operating System}
    \acro{RES}[RES]{Renewable Energy Source}
    \acro{RI}[RI]{Research Infrastructure}
    \acro{RSE}[RSE]{Ricerca sul Sistema Energetico}
    \acro{SINTEF}[SINTEF]{Stiftelsen for industriell og teknisk forskning}
    \acro{SoC}[SoC]{State-of-Charge}
    \acro{TDSO}[TDSO]{Transmission and Distribution System Operator}
    \acro{TUD}[TUD]{Technische Universiteit Delft}
    \acro{uAPI}[uAPI]{universal API}
    \acro{wrt}[w.r.t.]{with respect to}
\end{acronym}



\begin{abstract}
    This paper investigates the integration and validation of multi-energy systems within the H2020 ERIGrid 2.0 project, focusing on the deployment of the JaNDER software middleware and~\acf{uAPI} to establish a robust, high-data-rate, and low-latency communication link between~\acfp{RI}. The middleware facilitates seamless integration of~\acp{RI} through specifically designed transport layers, while the~\ac{uAPI} provides a simplified and standardized interface to ease deployment. A motivating case study explores the provision of power-to-heat services in a local multi-energy district, involving laboratories in Denmark, Greece, Italy, the Netherlands, and Norway, and analyzing their impact on electrical and thermal networks. This paper not only demonstrates the practical application of~\acl{GDS} and~\acl{HIL} technologies but also highlights their effectiveness in enhancing system flexibility and managing grid dynamics under various operational scenarios.
\end{abstract}










\section{Introduction}
\label{sec:introduction}

In the evolving landscape of multi-energy systems, the drive towards integrating diverse energy sources and optimizing their operations remains a pressing concern~\cite{MancarellaEnergy2014, LiuTranInduInfor2022}. This integration is crucial as it encompasses various dimensions of energy systems including power, heat, and transport, all undergoing significant transformations due to decentralization, decarbonization, and digitalization~\cite{SorknaesAppliedEnergy2015, LinAppliedThermal2024}. Traditionally, research in this domain has been skewed towards optimizing the planning and operational strategies within isolated engineering frameworks, often overlooking the intricate interconnections and interdependencies that exist among different energy networks~\cite{LundEnergy2018, Zhang2017TSG}.

Recent advancements have triggered a wave of innovation aimed at addressing these complexities, with significant emphasis on the development of new tools and methodologies that span across different engineering domains~\cite{WidlOSMSES2022, Gehrke2023Asia}. Notably, the concept of \ac{HIL} and its derivative, \ac{GDS}, have emerged as pivotal in enhancing the validation and deployment of novel multi-energy systems~\cite{Monti2018IEEEPEM, Syed2023IEEETPS}. These methodologies facilitate accelerated validation processes and enable the harnessing of distributed resources and expertise, thereby broadening the scope of what can be achieved within the constraints of individual \acp{RI}.

This paper explores the intersection of technological advancements and their practical applications, particularly emphasizing the use of \ac{HIL} and \ac{GDS} in multi-energy systems. Specifically, within the framework of the H2020 ERIGrid 2.0 project\footnote{\url{https://erigrid2.eu/}}, it examines the deployment of the JaNDER~\cite{Pellegrino2020Chapter, PellegrinoEnergies2020} software middleware and \acf{uAPI}~\cite{Rajkumar2024MSCPES, Gehrke2023Asia} for establishing a secure, reliable, high-data-rate, and low-latency communication between~\acfp{RI}. 
A motivating case study focuses on providing ancillary services through power-to-heat strategies in a local multi-energy district, examining their impact on electrical and thermal networks. This initiative aims to enhance power system flexibility as requested by the~\aclp{TDSO}, utilizing a combination of electric and thermal storage systems, alongside demand response strategies from controllable loads such as heat pumps and electric boilers~\cite{RodioAEIT2020}. Additionally, the heating system is designed to offer flexibility to the electrical system, e.g. for congestion management and power balancing services. Consequently, the key contributions of this paper are:
\begin{itemize}
    \item Providing a detailed description of the software setup used for implementing communication and conducting geographically distributed experiments among~\acp{RI} using the JaNDER open-source middleware and the \ac{uAPI}. The presented case study serves as a practical guide that other researchers can use, modify, or integrate to test their algorithms and understand different approaches within their~\acp{RI}. This enables the study of performance and stability in multi-energy systems.
    \item Demonstrating the feasibility of providing power-to-heat services in a local multi-energy district and assessing their impacts on electrical and thermal networks.  Furthermore, it highlights the potential of such services to manage congestions and provide balancing power.
\end{itemize}

The remainder of this paper is organized as follows. Section~\ref{sec:motivatingCaseStudy} discusses the motivating case study. Section~\ref{sec:distributedRISetup} details the software framework for~\ac{RI} communication and experiment management. The experimental results of~\ac{GDS} are presented in Section~\ref{sec:experimentalResults}. Section~\ref{sec:conclusions} concludes the paper.



\section{Motivating Case Study}
\label{sec:motivatingCaseStudy}

\begin{figure*}[tb]
    \centering
    \scalebox{0.75}{
    \begin{tikzpicture}
        \node (BESS) at (-3, 0) [draw, rectangle, minimum width=3em, minimum height=3em, text centered, text width=6em, font=\footnotesize, line width=0.5pt]{BESS\\ \textbf{[SINTEF]}};
        \node (GridForming-SINTEF) at (0, 0) [draw, rectangle, minimum width=3em, minimum height=3em, text centered, text width=6em, font=\footnotesize, line width=0.5pt]{Grid Forming\\ Converter \textbf{[SINTEF]}};
        \node (DistributionGrid) at (3, 0) [draw, rectangle, minimum width=3em, minimum height=3em, text centered, text width=6em, font=\footnotesize, line width=0.5pt]{Distribution Grid (RTDS)\\ \textbf{[TUD]}};
        \node (GridForming-RSE) at (6, 0) [draw, rectangle, minimum width=3em, minimum height=3em, text centered, text width=6em, font=\footnotesize, line width=0.5pt]{Grid Forming\\ Converter\\ \textbf{[RSE]}};
        \node (ElectricalNetwork) at (9, 1.75) [draw, rectangle, minimum width=3em, minimum height=3em, text centered, text width=6em, font=\footnotesize, line width=0.5pt]{Electrical Network\\ \textbf{[RSE]}};
        \node (CHP) at (12, 1.75) [draw, rectangle, minimum width=3em, minimum height=3em, text centered, text width=6em, font=\footnotesize, line width=0.5pt]{CHP\\ \textbf{[RSE]}};
        \node (GridFollowing-RSE) at (9, 0) [draw, rectangle, minimum width=3em, minimum height=3em, text centered, text width=6em, font=\footnotesize, line width=0.5pt]{Grid Following\\ Converter\\ \textbf{[RSE]}};
        \node (EHP) at (9, -1.75) [draw, rectangle, minimum width=3em, minimum height=3em, text centered, text width=6em, font=\footnotesize, line width=0.5pt]{EHP\\ \textbf{[CRES]}};
        \node (ThermalCRES) at (9, -3.5) [draw, rectangle, minimum width=3em, minimum height=3em, text centered, text width=6em, font=\footnotesize, line width=0.5pt]{Thermal Network\\ \textbf{[CRES]}};.
        \node (CouplingUnit) at (12, -1.75) [draw, rectangle, minimum width=3em, minimum height=3em, text centered, text width=6em, font=\footnotesize, line width=0.5pt]{Coupling Unit\\ \textbf{[DTU]}};
        \node (HeatDistribution) at (12, -3.5) [draw, rectangle, minimum width=3em, minimum height=3em, text centered, text width=6em, font=\footnotesize, line width=0.5pt]{District Heating Network\\ \textbf{[DTU]}};
        \node (Controller) at (9, 3.75) [draw, rectangle, minimum width=3em, minimum height=3em, text centered, text width=6em, font=\footnotesize, line width=0.5pt]{Centralized Supervisory\\ Controller \textbf{[RSE]}};

        \draw[-latex, blue!50, line width=0.75pt] (BESS) -- (GridForming-SINTEF);
        \draw[-latex, blue!50, line width=0.75pt] (GridForming-SINTEF.10) -- node [above, yshift=0.75em, text centered, font=\scriptsize, text width=5em]{$P_\mathrm{el_{SIN}}$,\\ $Q_\mathrm{el_{SIN}}$} (DistributionGrid.170);
        \draw[latex-, blue!50, line width=0.75pt] (GridForming-SINTEF.-10) -- node [below, yshift=-0.5em, xshift=0.1em, text centered, font=\scriptsize, text width=5em]{$V^\mathrm{ref}_\mathrm{SIN}$,\\ $f^\mathrm{ref}_\mathrm{SIN}$} (DistributionGrid.190);
        \draw[-latex, blue!50, line width=0.75pt] (DistributionGrid.-10) -- node [below, yshift=-0.5em, xshift=0.1em, text centered, font=\scriptsize]{$V^\mathrm{ref}_\mathrm{RSE}$, $f^\mathrm{ref}_\mathrm{RSE}$} (GridForming-RSE.190);   
        \draw[latex-, blue!50, line width=0.75pt] (DistributionGrid.10) -- node [above, yshift=0.75em, text centered, font=\scriptsize, text width=5em]{$P_\mathrm{el_{RSE}}$,\\ $Q_\mathrm{el_{RSE}}$} (GridForming-RSE.170); 
        \draw[-latex, blue!50, line width=0.75pt] (GridForming-RSE) |- (ElectricalNetwork);
        \draw[-latex, blue!50, line width=0.75pt] (ElectricalNetwork) -- (CHP);
        \draw[-latex, blue!50, line width=0.75pt] (ElectricalNetwork) -- (GridFollowing-RSE); 
        \draw[-latex, blue!50, line width=0.75pt] (CHP) node[right, yshift=-5em, font=\scriptsize]{$P_\mathrm{th_{CHP}}$} -- (CouplingUnit);
        \draw[-latex, blue!50, line width=0.75pt] (EHP) -- (GridFollowing-RSE);
        \draw[-latex, red!75, line width=0.75pt] (EHP) -- node[right, font=\scriptsize]{$P_\mathrm{th_{CRES}}$} (ThermalCRES);
        \draw[-latex, red!75, line width=0.75pt] (CouplingUnit) -- node[right, font=\scriptsize]{$\bar{P}_\mathrm{DTU}$} (HeatDistribution);
        \draw[-latex, line width=0.75pt] (HeatDistribution.south) -- ($(HeatDistribution.south) - (0,0.5)$) -- ($(HeatDistribution.south) - (5,0.5)$) -- ($(HeatDistribution.south) - (5,0)$) -- ($(HeatDistribution.south) - (6,0)$) node[right, xshift=-2.75em, font=\scriptsize, text width=5em]{$T_\mathrm{DTU}$, \\ $T_\mathrm{{CRES}}$};
        \draw[-, line width=0.75pt] (ThermalCRES.west) -- ($(ThermalCRES.west) - (0.825,0)$) -- ($(HeatDistribution.south) - (5,0)$);
        \draw[-latex, myGreen, line width=0.75pt] ($(BESS.north) + (0,0.5)$) node[right, text centered, font=\scriptsize]{$P_\mathrm{el_{SIN}}^\mathrm{ref}$} -- (BESS);
        \draw[-latex, myGreen, line width=0.75pt] (BESS.south) -- ($(BESS.south) - (0,0.5)$) node[right, text centered, font=\scriptsize]{$\mathrm{SoC}$};
        \draw[-latex, line width=0.75pt] (DistributionGrid.north) -- node[above, text centered, font=\scriptsize, yshift=0.75em, xshift=-0.5em, text width=5em]{$P_\mathrm{el_{SIN}}, \dots, P_\mathrm{el_{RSE}}$\\ $Q_\mathrm{el_{SIN}}, \dots, Q_\mathrm{el_{RSE}}$} ($(DistributionGrid.north) + (0,0.5)$);
        \draw[-latex, myGreen, line width=0.75pt]  ($(CHP.north) + (0,0.5)$) node[right, text centered, text width=5em, xshift=-1em, font=\scriptsize]{$P_\mathrm{el_{RSE}}^\mathrm{ref}$} -- (CHP.north);
        \draw[-latex, myGreen, line width=0.75pt]  ($(EHP.west) - (0.5,0)$) node[above left, text centered, text width=5em, font=\scriptsize, xshift=1.75em]{$\mathrm{ON/OFF}$} -- (EHP.west);
        \draw[-latex, myGreen, line width=0.75pt] ($(Controller.west) - (0.5,0)$) node[left, text centered, font=\scriptsize]{$\mathrm{SoC}$} -- (Controller.west);
        \draw[-latex, myGreen, line width=0.75pt] (Controller.10) -- node[right, xshift=0.5em, text centered, yshift=0.5em, font=\scriptsize]{$\mathrm{ON/OFF}$} ($(Controller.10) + (0.5,0)$);
        \draw[-latex, myGreen, line width=0.75pt] (Controller.-10) -- node[right, font=\scriptsize, xshift=0.5em, text width=10em]{$P_\mathrm{el_{SIN}}^\mathrm{ref}$, $Q_\mathrm{el_{SIN}}^\mathrm{ref}$,\\ $P_\mathrm{el_{RSE}}^\mathrm{ref}$, $Q_\mathrm{el_{SIN}}^\mathrm{ref}$} ($(Controller.-10) + (0.5,0)$);

        \draw (0,-2.4) node[text centered, font=\footnotesize]{\textsc{Legend}};
        \draw[-latex, line width=0.75pt] (-2, -2.9) -- (-1, -2.9) node[right, text centered, font=\footnotesize]{Measurement signal};
        \draw[-latex, blue!50, line width=0.75pt] (-2, -3.4) -- (-1, -3.4) node[right, text centered, blue!50, font=\footnotesize]{Electrical signal};
        \draw[-latex, red!75, line width=0.75pt] (-2, -3.9) -- (-1, -3.9) node[right, text centered, red!75, font=\footnotesize]{Thermal connection};
        \draw[-latex, myGreen, line width=0.75pt] (-2, -4.4) -- (-1, -4.4) node[right, text centered, myGreen, font=\footnotesize]{Control connection};
        \draw[dashed] (2,-2.05) -- (2,-4.9) -- (-2.25,-4.9) -- (-2.25,-2.05) -- cycle; 

        \node (Norway-flag) at (-3,2.9) [text centered]{\includegraphics[scale=0.075]{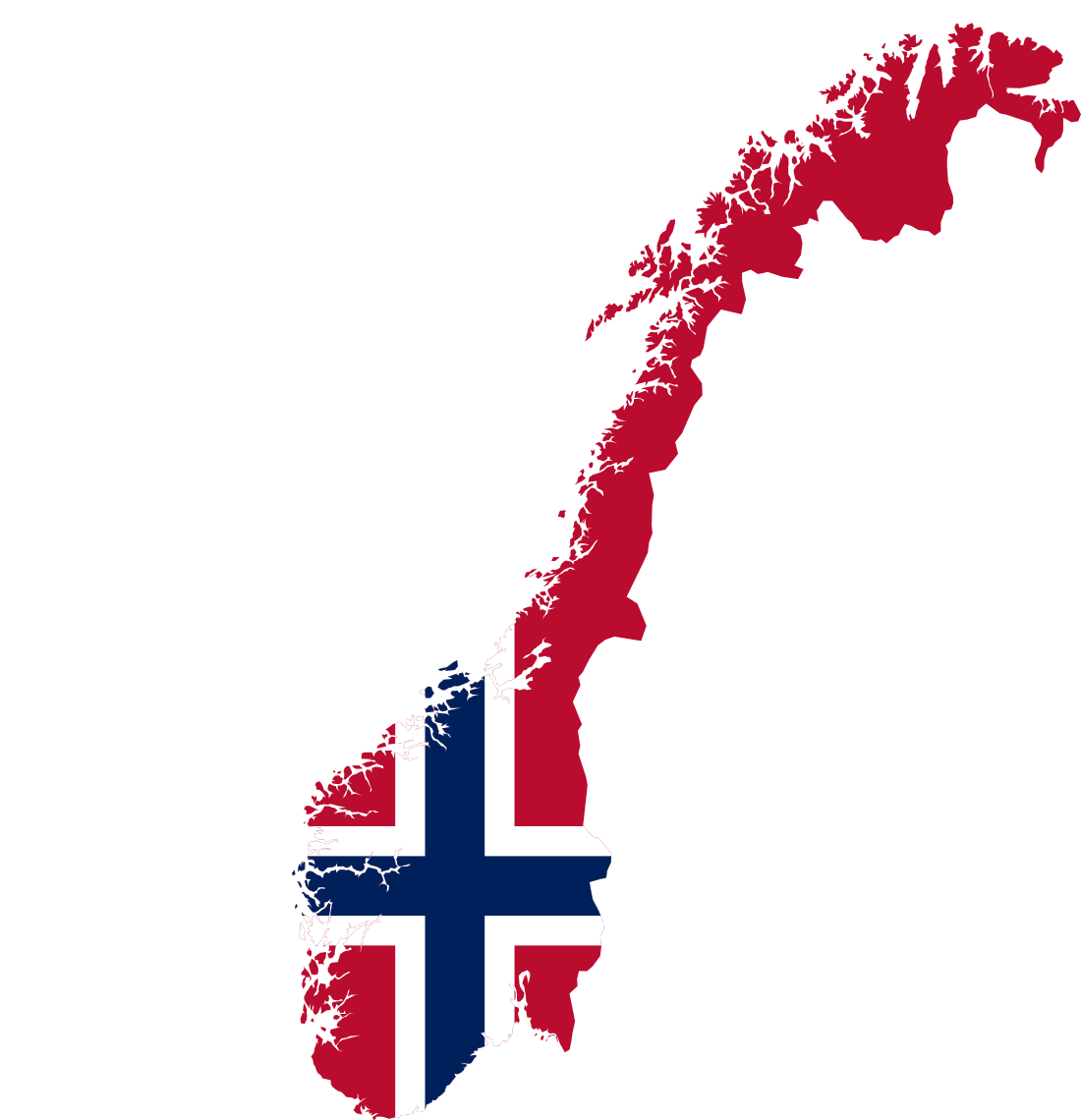}};
        \node (Denmark-flag) at (14.35,-3) [text centered]{\includegraphics[scale=0.075]{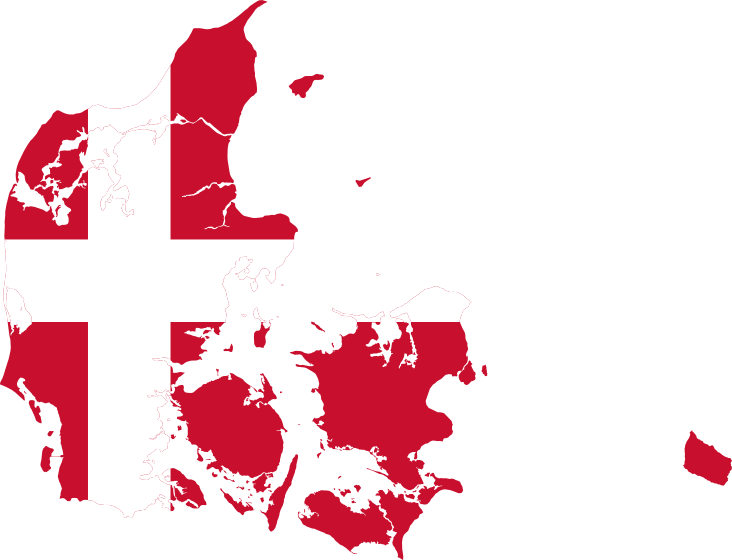}};
        \node (Italy-flag) at (14.35,3.0) [text centered]{\includegraphics[scale=0.065]{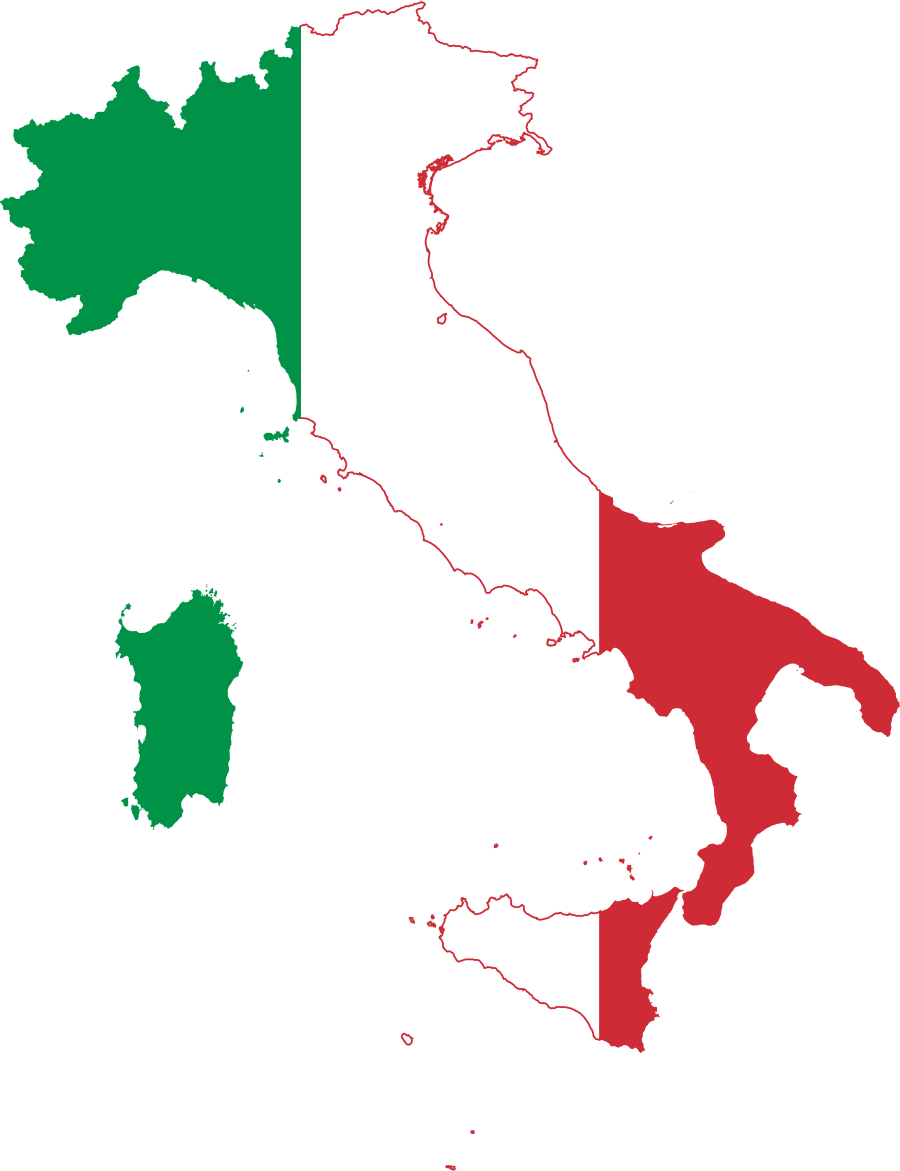}};
        \node (Greece-flag) at (6,-2.5) [text centered]{\includegraphics[scale=0.125]{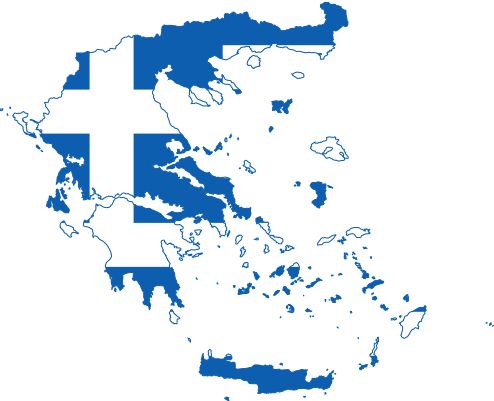}};
        \node (Netherlands-flag) at (3,3) [text centered]{\includegraphics[scale=0.10]{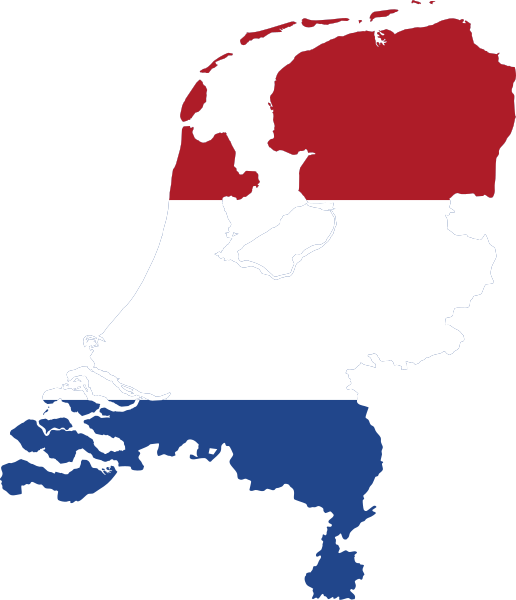}};

        \draw[densely dotted] (-4.5, 4.5) -- (1.5,4.5) -- (1.5,-1.25) -- (-4.5,-1.25) -- cycle; 
        \draw[densely dotted] (1.6,4.5) -- (1.6,-1.25) -- (4.5,-1.25) -- (4.5,4.5) -- cycle; 
        \draw[densely dotted] (4.6,-0.85) -- (4.6,4.5) -- (15.75,4.5) -- (15.75,-0.85) -- cycle; 
        \draw[densely dotted] (15.75,-0.95) -- (10.6,-0.95) -- (10.6,-5.0) -- (15.75,-5.0) -- cycle; 
        \draw[densely dotted] (4.60,-0.95) -- (10.5,-0.95) -- (10.5,-5.0) -- (4.60,-5.0) -- cycle; 
        
        \draw[red!75, dashed, line width=1.25pt, loop] (4.6,-0.95) -- (4.6,-5.0) -- (15.75,-5.0) -- (15.75,-0.95) -- cycle;
        \draw[blue!50, dashed, line width=1.25pt] (15.75, 4.5) -- (-4.5, 4.5) -- (-4.5,-1.25) -- (4.5,-1.25) -- (4.5,-0.85) -- (10.6,-0.85) -- (15.75, -0.85) -- cycle;
    \end{tikzpicture}
    }
    \vspace*{-0.50em}
    \caption{Schematic representation of the multi-energy district case study, with the electrical and thermal subsystems highlighted by blue and red dashed boxes, respectively.}
    \label{fig:demo4}
\end{figure*}
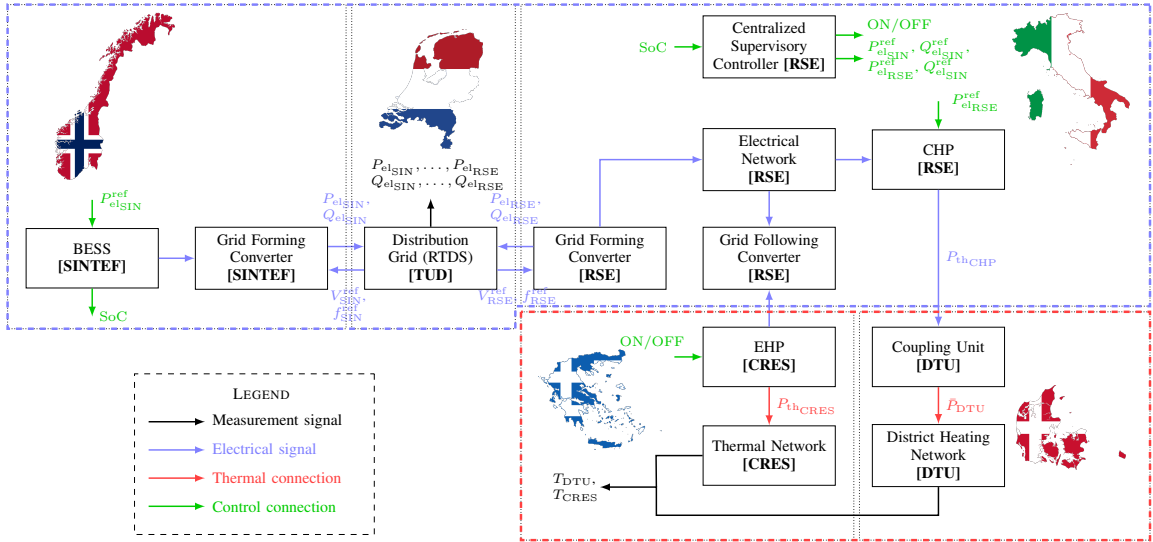

The case study considered here explores the provision of critical services to the electrical grid, with particular emphasis on congestion management -- including the constraints associated with electrical import and export -- and regulating power provision. The focal point of this case study is the assessment of a \ac{CSC} within a geographically distributed local multi-energy system. This analysis is crucial for understanding the operational dynamics of system components interconnected across various regions.

\begin{figure}[tb]
    \centering
    \scalebox{0.7}{
    \begin{tikzpicture}
    
    \draw[line width=1.25pt] (0,-1.5) -- (0,1.5);
    \draw[line width=1.25pt] (0,0) -- (1,0);
    \node at (1.4,0) [draw, circle, minimum size=2em, line width=1.25pt]{}; 
    \node at (1.8,0) [draw, circle, minimum size=2em, line width=1.25pt]{};
    \draw[line width=1.25pt] (2.2,0) -- (3,0);
    \draw[line width=1.25pt] (3,-2) -- (3,2);
    \draw[line width=1.25pt, latex-] (1.0,0.75) -- (1.6,-0.75);

    \node at (0.2,1.15) [text centered, rotate=90]{\footnotesize 20kV};
    \node at (3.2,1.6) [text centered, rotate=90]{\footnotesize 0.4kV};

    \draw[line width=1.25pt] (3,0) -- (11,0);
    \draw[line width=1.25pt] (5,0) -- (5,-1);
    \draw[line width=1.25pt] (7,0) -- (7,-1);
    \draw[line width=1.25pt] (7,0) -- (7,1);
    \draw[line width=1.25pt] (9,0) -- (9,-1);
    \draw[line width=1.25pt] (10,0) -- (10,1);
    \draw[line width=1.25pt] (11,0) -- (11,-1);

    \node at (4,0) [draw, rectangle, minimum size=0.1em, fill=white, line width=1.25pt]{};
    \node at (5,-0.5) [draw, rectangle, minimum size=0.1em, fill=white, line width=1.25pt]{};
    \node at (6,0) [draw, rectangle, minimum size=0.1em, fill=white, line width=1.25pt]{};
    \node at (7,-0.5) [draw, rectangle, minimum size=0.1em, fill=white, line width=1.25pt]{};
    \node at (7,0.5) [draw, rectangle, minimum size=0.1em, fill=white, line width=1.25pt]{};
    \node at (8,0) [draw, rectangle, minimum size=0.1em, fill=white, line width=1.25pt]{};
    \node at (9,-0.5) [draw, rectangle, minimum size=0.1em, fill=white, line width=1.25pt]{};
    \node at (9.5,0) [draw, rectangle, minimum size=0.1em, fill=white, line width=1.25pt]{};
    \node at (10,0.5) [draw, rectangle, minimum size=0.1em, fill=white, line width=1.25pt]{};
    \node at (10.5,0) [draw, rectangle, minimum size=0.1em, fill=white, line width=1.25pt]{};
    \node at (11,-0.5) [draw, rectangle, minimum size=0.1em, fill=white, line width=1.25pt]{};

    \draw[line width=1.25pt] (4.5,-1) -- (5.5,-1);
    \draw[line width=1.25pt] (6.5,-1) -- (7.5,-1);
    \draw[line width=1.25pt] (6.5,1) -- (7.95,1);
    \draw[line width=1.25pt] (8.05,-1) -- (9.5,-1);
    \draw[line width=1.25pt] (9.5,1) -- (10.95,1);
    \draw[line width=1.25pt] (10.5,-1) -- (11.95,-1);

    \draw[-latex, line width=1.25pt] (4.75,-1) -- (4.75, -1.5);
    \draw[-latex, line width=1.25pt] (7.25,1) -- (7.25, 1.5);
    \draw[-latex, line width=1.25pt] (7.75,1) -- (7.75, 1.5);
    \draw[-latex, line width=1.25pt] (8.25,-1) -- (8.25, -1.5);
    \draw[-latex, line width=1.25pt] (8.75,-1) -- (8.75, -1.5);
    \draw[-latex, line width=1.25pt] (10.25,1) -- (10.25, 1.5);
    \draw[-latex, line width=1.25pt] (10.75,1) -- (10.75, 1.5);
    \draw[-latex, line width=1.25pt] (11.25,-1) -- (11.25, -1.5);
    \draw[-latex, line width=1.25pt] (11.75,-1) -- (11.75, -1.5);

    \node at (7.40, 1.15) [draw, circle, scale=0.2, fill=black, line width=1.25pt]{};
    \node at (7.60, 1.15) [draw, circle, scale=0.2, fill=black, line width=1.25pt]{};
    \node at (8.40,-1.15) [draw, circle, scale=0.2, fill=black, line width=1.25pt]{};
    \node at (8.60,-1.15) [draw, circle, scale=0.2, fill=black, line width=1.25pt]{};
    \node at (10.40, 1.15) [draw, circle, scale=0.2, fill=black, line width=1.25pt]{};
    \node at (10.60, 1.15) [draw, circle, scale=0.2, fill=black, line width=1.25pt]{};
    \node at (11.40,-1.15) [draw, circle, scale=0.2, fill=black, line width=1.25pt]{};
    \node at (11.60,-1.15) [draw, circle, scale=0.2, fill=black, line width=1.25pt]{};

    \draw[draw=black, line width=1.25pt] (6.75,-1.65) rectangle ++(0.5,0.5);
    \draw[draw=black, line width=1.25pt] (10.2,-1.65) rectangle ++(0.5,0.5);
    \draw[draw=black, line width=1.25pt] (9.2, 1.15) rectangle ++(0.5,0.5);
    \draw[draw=black, line width=1.25pt] (6.2, 1.15) rectangle ++(0.5,0.5);
    \draw[draw=black, line width=1.25pt] (9.2,-1.65) rectangle ++(0.5,0.5);
    \draw[line width=1.25pt] (7.0, -1.15) -- (7.0, -0.98);
    \draw[line width=1.25pt] (6.5, 1.15) -- (6.5, 0.98);
    \draw[line width=1.25pt] (9.5, 1.15) -- (9.5, 0.98);
    \draw[line width=1.25pt] (9.5, -1.15) -- (9.5, -0.98);
    \draw[line width=1.25pt] (10.5, -1.15) -- (10.5, -0.98);
    \draw[line width=1.25pt] (6.75,-1.65) -- (7.25,-1.15);
    \draw[line width=1.25pt] (10.20,-1.65) -- (10.70,-1.15);
    \draw[line width=1.25pt] (9.70,-1.15) -- (9.20,-1.65);
    \draw[line width=1.25pt] (9.20,1.15) -- (9.70,1.65);
    \draw[line width=1.25pt] (6.20,1.15) -- (6.70,1.65);

    \node at (6.52,1.265) [text centered]{\footnotesize $\sim$};
    \node at (6.35,1.475) [text centered]{\footnotesize $=$};
    \node at (9.52,1.265) [text centered]{\footnotesize $\sim$};
    \node at (9.35,1.475) [text centered]{\footnotesize $=$};
    \node at (9.35,-1.325) [text centered]{\footnotesize $=$};
    \node at (9.52,-1.525) [text centered]{\footnotesize $\sim$};
    \node at (10.35,-1.325) [text centered]{\footnotesize $=$};
    \node at (10.52,-1.525) [text centered]{\footnotesize $\sim$};
    \node at (6.90,-1.325) [text centered]{\footnotesize $=$};
    \node at (7.075,-1.525) [text centered]{\footnotesize $\sim$};

    \node at (4.75, -1.75) [text centered]{\footnotesize Load 5};
    \node at (7.5,  1.75) [text centered]{\footnotesize Load 4};
    \node at (8.5, -1.75) [text centered]{\footnotesize Load 3};
    \node at (10.5,  1.75) [text centered]{\footnotesize Load 2};
    \node at (11.5,  -1.75) [text centered]{\footnotesize Load 1};

    \node at (7.0,  -1.85) [text centered]{\footnotesize BESS};
    \node at (9.5, -1.85) [text centered]{\footnotesize PV 3};
    \node at (10.5,  -1.85) [text centered]{\footnotesize PV 1};
    \node at (9.45,  1.85) [text centered]{\footnotesize PV 2};
    \node at (6.45,  1.85) [text centered]{\footnotesize PV 4};

    \draw[draw=black] (4.25,-2.05) rectangle ++(1.4,1.2);
    \draw[draw=black] (6.25,-2.05) rectangle ++(1.4,1.2);
    \draw[draw=black] (7.95,-2.05) rectangle ++(1.95,1.2);
    \draw[draw=black] (10.05,-2.05) rectangle ++(1.95,1.2);

    \node at (5.20, -2.25) [text centered]{\footnotesize PCC1};
    \node at (7.20, -2.25) [text centered]{\footnotesize PCC2};
    \node at (9.50, -2.25) [text centered]{\footnotesize PCC3};
    \node at (11.60,-2.25) [text centered]{\footnotesize PCC4};
    
    \end{tikzpicture}
    }
    \vspace*{-1.6em}
    \caption{Schematic representation of the CIGRE LV-distribution benchmark grid along with the~\acs{PCC}.}
    \label{fig:CIGRENetwork}
\end{figure}
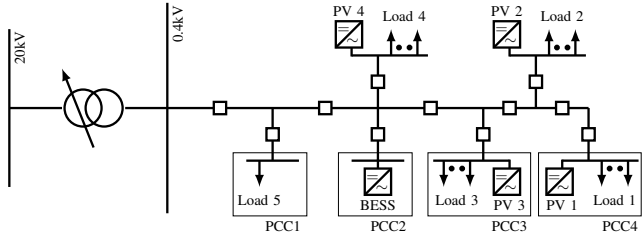

The experimental setup encompasses a tripartite system: an electrical grid with energy sources, a thermal network, and an advanced control system, all graphically represented in Figure~\ref{fig:demo4}. The \textit{thermal infrastructure} (red dashed box) is divided into two subsystems. The first is a~\ac{DHN} located in Denmark at the~\ac{DTU}, interfaced with a~\ac{CHP} unit based in Italy at the \ac{RSE}, connected through an electrical coupling unit. The second subsystem consists of a thermal load (L1) powered by an~\ac{EHP}, both situated in Greece at the~\ac{CRES}. The \textit{electrical infrastructure} (blue dashed box) includes a distribution grid and a few controllable units. The distribution grid adopts the CIGRE LV-distribution benchmark~\cite{ManiatopoulosICCS2017}, characterized by a $\SI{0.4}{\kilo\volt}$, $\SI{50}{\hertz}$ low-voltage system with resources working as controllable current sources, allowing precise control over active ($P$) and reactive power ($Q$) through the calculation of current magnitude and phase relative to the bus voltage. The schematic representation of this network is detailed in Figure~\ref{fig:CIGRENetwork}. This network is emulated through an RTDS real-time simulator\footnote{\url{https://www.rtds.com/}} located in the Netherlands at the~\ac{TUD}, with two physically different nodes virtually interfaced in a \ac{GDS} setup. To these nodes, \acs{PCC}2 and \acs{PCC}4, are connected the simulated \ac{BESS} of \ac{SINTEF},  and the \ac{CHP} with the \ac{EHP}, respectively. The \textit{control system} delves with accurately determining setpoints for these controllable entities to ensure the delivery of required services. 

\begin{figure}
    \centering
    \includegraphics[width=0.95\columnwidth]{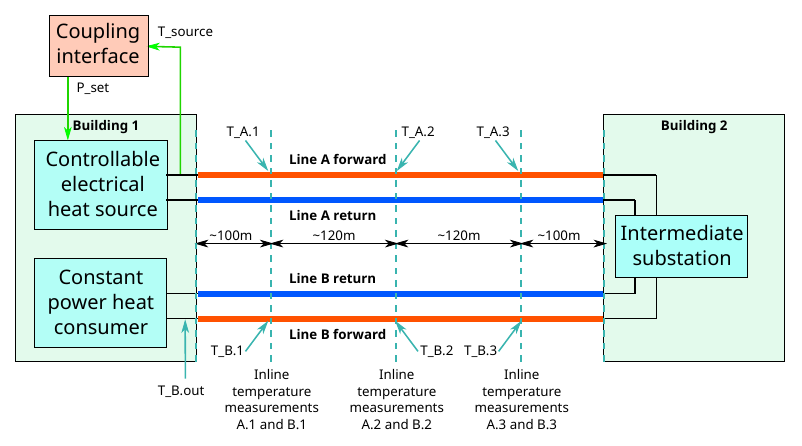}
    \vspace*{-0.70em}
    \caption{Schematic representation of the \ac{DHN} at \ac{DTU}, with approximate  measurement locations marked}
    \label{fig:dtu_heatlayout}
\end{figure}

Regarding the hardware setup of the experiment, it includes several key components. The~\ac{EHP} (\acs{CRES}) has nominal power $\SI{16}{\kilo\watt}$ and capacity $\SI{18}{\kilo\volt\ampere}$, with controllable variables including its operating state (ON/OFF). Various measurements are recorded, such as active, reactive, and apparent power, energy consumption, voltage, frequency, and indoor temperature at multiple points. Interfacing with other~\acp{RI} involves data collection via an eWon 4001 datalogger and control through a Raspberry Pi 4 Model B. 
The \ac{DHN} (\acs{DTU}) consists of two \SI{440}{\meter} long double pipes connecting two buildings as depicted in Figure~\ref{fig:dtu_heatlayout}. Building 1 houses a controllable electrical heat source and heat consumer, while Building 2 functions as a thermal substation interconnecting the pipes, creating a total pipe length of \SI{880}{\meter} between the source and sink. The heat source includes nine electrical flow heaters totaling \SI{22.5}{\kilo\watt}, feeding into a \SI{200}{\liter} accumulator tank. To address the size mismatch between the \ac{CHP} plant at \acs{RSE} (\SIrange{46}{81}{\kilo\watt}) and the \acs{DTU} heat source (\SIrange{0}{22.5}{\kilo\watt}), a linear mapping function offsets and scales the thermal power setpoints received through the coupling interface, ensuring the full controllable range of the \ac{CHP} plant is utilized. 
The three-phase distribution grid (\acs{TUD}) operates at $\SI{0.4}{\kilo\volt}$, $\SI{50}{\hertz}$ and supplies five loads (see Figure~\ref{fig:CIGRENetwork}). The~\ac{BESS} managed by \acs{SINTEF} includes a converter and a power amplifier, along with measurement points including~\ac{SoC} and instantaneous power. Lastly, the~\ac{CHP} (\acs{RSE}) plant utilizes a natural gas internal combustion engine, bidirectional converters, and power amplifiers within a three-phase low-voltage grid setup. 

\section{Distributed RI Setup}
\label{sec:distributedRISetup}

This section details the software framework developed to facilitate communication among the~\acp{RI} engaged in the experimental setups of the motivating case study. At the core of this framework is the deployment of~\ac{uAPI}~\cite{Rajkumar2024MSCPES, Gehrke2023Asia}, powered by the JaNDER middleware~\cite{Pellegrino2020Chapter, PellegrinoEnergies2020}. The \ac{uAPI} serves as a transport-independent abstraction layer, enabling the use of various software middleware for multi-\ac{RI} experiments. This eliminates the need to implement individual laboratory interfaces and provides common core functionality, such as accessing a list of available signals, \ac{RI} statuses, and more. JaNDER complements the \ac{uAPI} by enabling secure and efficient data exchange between~\acp{RI} via an \texttt{HTTPS}-secured internet connection. Its primary functionality includes replicating infrastructure data across network nodes, ensuring that data captured in local database is simultaneously mirrored in a cloud-based database. This setup guarantees data consistency and redundancy without altering or interpreting the data's structure. 

\begin{figure}
    \centering
    \includegraphics[width=0.90\columnwidth]{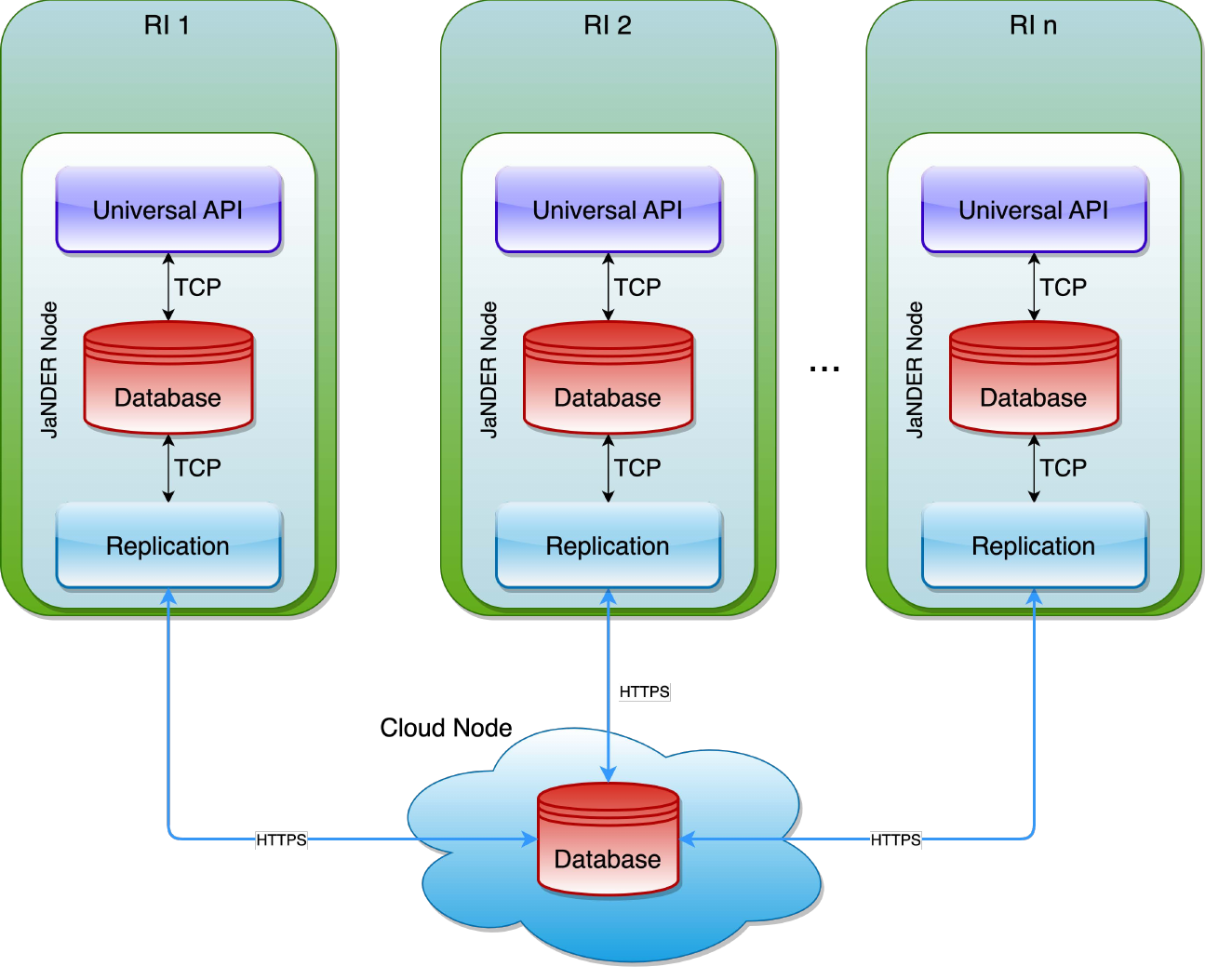}
    \vspace*{-0.65em}
    \caption{Diagram of the JaNDER middleware, featuring~\acp{RI} each equipped with the~\ac{uAPI} and local database instances, which are connected to a shared database in the cloud node.}
    \label{fig:janderScheme}
\end{figure}

A schematic representation of the JaNDER's software architecture can be found in Figure~\ref{fig:janderScheme}, and the source code used for the experiments is publicly available on GitHub\footnote{\url{https://github.com/ERIGrid2/JRA-3.1-JaNDER-API}}. These resources provide a practical guide for other researchers to replicate, modify, or integrate the setup for testing their algorithms and studying different approaches within their \acp{RI}. This framework enables the study of performance and stability in multi-energy systems, offering a robust platform for real-time experimentation and data analysis.
Further details about the open-source middleware are discussed in~\cite{Pellegrino2020Chapter, PellegrinoEnergies2020} but are omitted here for brevity.

The operational deployment of the JaNDER middleware and \ac{uAPI} requires a Linux-based \ac{OS}, such as Ubuntu. This choice promotes the use of open-source tools and avoids dependency on specific proprietary licenses. The setup process involves several key steps. Firstly, a Linux-based \ac{OS} is installed on a dedicated or secondary machine separate from the \ac{RI}'s main data monitoring and control infrastructure. Following the \ac{OS} installation, Docker\footnote{\url{https://www.docker.com/}} is used to containerize applications, ensuring secure, isolated, and portable environments. This streamlines the software deployment process and minimizes configuration requirements. Secure communication between~\acp{RI} is facilitated by generating private and public certificates. These certificates provide robust authentication and encryption mechanisms~\cite{Pellegrino2020Chapter, Gehrke2023Asia}. Finally, environment configuration involves updating configuration files to include namespaces for the \acp{RI}, such as \ac{RSE}, \ac{TUD}, and \ac{CRES}, ensuring correct identification within the network. 

Once the development environment is configured, the Docker container must be compiled. 
At this stage, communication between the \acp{RI} is established, facilitating data exchange via the \ac{uAPI}. The \ac{uAPI} serves as an abstraction layer that enables integration with SCADA systems by defining a set of REST functions~\cite{Rajkumar2024MSCPES}. The implementation of these functions supports \texttt{GET/SET} operations, which are essential for real-time monitoring and control of the multi-energy system.  For instance, the \ac{uAPI} defines endpoints for data retrieval (\texttt{GET}) and data updates (\texttt{SET}). 
The REST functions allow the SCADA system to request current data states or send control commands to the various components of the multi-energy system. Detailed documentation and code examples to implement these REST functionalities are available in the \ac{uAPI} repository\footnote{\url{https://github.com/ERIGrid2/JRA-3.1-api}}. 




\section{Experimental Results}
\label{sec:experimentalResults}

\begin{table}
    \caption{Signals exchanged among \acp{RI}, including their symbols and operational ranges.}
    \label{tab:hardware_signals}
    \vspace*{-0.65em}
    \centering
    \resizebox{\columnwidth}{!}{%
    \begin{tabular}{ c | c | c | c | c | c | c | c }
    \hline
    \textbf{Sym.} & \textbf{Unit} & \textbf{Min} & \textbf{Max} & \textbf{Sym.} & \textbf{Unit} & \textbf{Min} & \textbf{Max} \\ 
    \hline\hline
    $P_\mathrm{el_{SIN}}$ & $\si{\kilo\watt}$ & -40 & 40 & $Q_\mathrm{el_{SIN}}$ & $\si{\kilo\volt\ampere r}$ & -5 & 5\\ 
    $P_\mathrm{th_{CHP}}$ & $\si{\kilo\watt}$ & 46 & 81 & $P_\mathrm{el_{SIN}}^\mathrm{ref}$ & $\si{\kilo\watt}$ & -40 & 40\\ 
    $\bar{P}_\mathrm{DTU}$ & $\si{\kilo\watt}$ & 0 & 25 & \ac{SoC} & \% & 0 & 100 \\ 
    $V_\mathrm{SIN}^\mathrm{ref}$ & V & 150 & 400 & $f_\mathrm{SIN}^\mathrm{ref}$ & $\si{\hertz}$ & 48 & 52 \\ 
    $P_\mathrm{el_{RSE}}$ & $\si{\kilo\watt}$ & -100 & 100 & $Q_\mathrm{el_{RSE}}$ & $\si{\kilo\volt\ampere r}$ & -50 & 50 \\
    $V_\mathrm{RSE}^\mathrm{ref}$ & V & 150 & 400 & $f_\mathrm{RSE}^\mathrm{ref}$ & $\si{\hertz}$ & 48 & 52 \\
    $P_\mathrm{th_{CRES}}$ & $\si{\kilo\watt}$ & 0 & 30 & $T_\mathrm{DTU}$  & $\si{\celsius}$ & 0 & 100 \\
    \hline
    \end{tabular}%
    }
\end{table}

To validate the effectiveness of the proposed experimental setup and demonstrate the feasibility of providing ancillary services through power-to-heat strategies in a local multi-energy district, a sector-coupling experimental demonstration inspired by~\cite{WidlOSMSES2022} was conducted.

The performed experiments demonstrated the impact of providing flexibility services on both the electrical and thermal networks. The amount of flexibility requested by the system operator is achieved through the control of electric and thermal units, such as storage systems, heat pumps, thermal loads, and electric boilers. The distributed laboratory setup, as shown in Figure~\ref{fig:demo4}, utilizes the ~\ac{uAPI} for data exchange. Given the ``slower'' dynamics typical of electro-thermal experiments, which operate on the scale of seconds, the data-exchange rate is set between \SIrange{1}{2}{\hertz}. Two distinct working scenarios were delineated:
\begin{itemize}
    \item \textbf{Case 1 -- Overvoltage Scenario}: this scenario addresses an overvoltage condition triggered by high \ac{PV} generation, either \ac{PV}1 and \ac{PV}3 located at \acs{PCC}4 and \acs{PCC}3, respectively, as depicted in Figure~\ref{fig:CIGRENetwork}. This condition is coupled with low electricity demand. It is noteworthy that the \ac{DHN} does not influence the outcomes in this scenario.
    \item \textbf{Case 2 -- Undervoltage Scenario}: this scenario examines an undervoltage condition resulting from reduced or nonexistent \ac{PV} generation and elevated electricity demand. It mirrors the overvoltage scenario (Case 1) with the critical distinction of employing negative threshold values for the operational parameters of the units. This condition necessitates the engagement of both the \ac{CHP} unit and thermal storage to ensure grid stability.
\end{itemize}

This comprehensive case study aims to shed light on the operational efficacy of a \ac{CSC} within a distributed multi-energy framework, highlighting its capacity to manage grid congestions and provide regulating power under a spectrum of operational conditions.

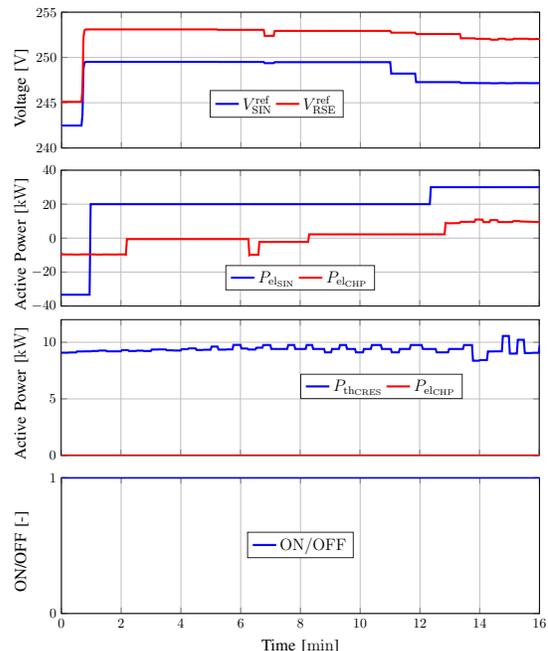
\begin{figure}[tb]
    \centering
    \input{matlabPlots/figure_4_overvoltage.tex}
    \vspace*{-0.65em}
    \caption{Reference voltage sent from the distribution grid (\acs{TUD}) to the grid forming converters (\acs{SINTEF} and \acs{RSE}), along with the active power generated by the \ac{BESS} and \ac{CHP} and enable signals of the \ac{EHP} in the overvoltage scenario.}
    \label{fig:overVoltage_1}
\end{figure}

Figure~\ref{fig:overVoltage_1} illustrates the outcomes of the overvoltage scenario aimed at addressing high \ac{PV} generation coupled with low electricity consumption, leading to an overvoltage condition. To prepare for this test, \ac{TUD} developed profiles for the \acp{PV} and simulated loads\footnote{The \acp{PV} and simulated loads profiles are not included here for brevity and as they are not crucial to the purpose of the case study which primarily focuses on showcasing the viability of deployment of the JaNDER software middleware and \ac{uAPI}.}, inducing a significant voltage rise at the nodes connected to \acs{SINTEF} ($V_\mathrm{SIN}^\mathrm{ref}$) and \acs{RSE} ($V_\mathrm{RSE}^\mathrm{ref}$). During the test, \acs{SINTEF} and \acs{RSE}, integrated into the simulation model (see Figure~\ref{fig:demo4}), active power consumption ($P_\mathrm{el_{SIN}}$ and $P_\mathrm{el_{RSE}}$) to mitigate voltage deviations. \acs{SINTEF} employed a simplified control approach, adjusting the active power setpoint of the battery in a stepwise manner, as depicted in the lower left plot of Figure~\ref{fig:overVoltage_1}. For instance, significant power consumption was activated if the voltage rise exceeded $\SI{5}{\%}$ and deactivated if it dropped below $\SI{0}{\%}$, ensuring stability during the test. Similarly, \acs{RSE} employed the same hysteresis approach to control the electricity consumption of \acs{CRES}' heat pump, reducing the voltage rise at the corresponding node, as shown in the plots on the right side of Figure~\ref{fig:overVoltage_1}. It is important to note that the $\SI{5}{\%}$ threshold is indicative and may vary based on the simulated profiles, highlighting the necessity for accurate estimations via offline simulations. Furthermore, in this scenario, where an increase in electric load is necessary, neither the thermal power of the combined heat and power (\ac{CHP}) unit nor the thermal capacity of the heat network (\ac{DTU}) can be utilized to mitigate the overvoltage condition.
 
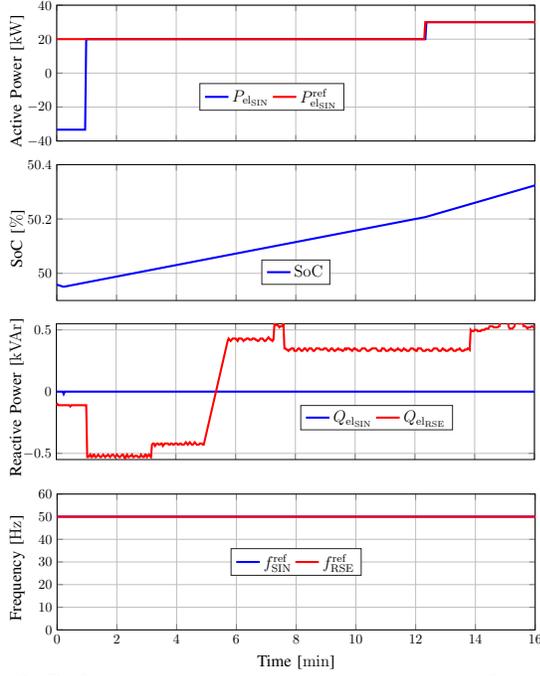
\begin{figure}[tb]
    \centering
    \input{matlabPlots/figure_5_overvoltage.tex}
    \vspace*{-0.65em}
    \caption{Reference and actual active power profiles at the \acs{SINTEF} facility, along with the \ac{SoC} of the \ac{BESS}, during the overvoltage scenario experiment. Additionally, reactive power and frequency data are included.}
    \label{fig:overVoltage_2}
\end{figure}

In this context, \acs{RSE} could still communicate the thermal ($P_\mathrm{th_{CHP}}$) setpoint value to \acs{DTU}, but it would need to be set to zero as it does not play any role. Whereas, Figure~\ref{fig:overVoltage_2} illustrates the reference ($P_\mathrm{el_{SIN}}^\mathrm{ref}$) and actual ($P_\mathrm{el_{SIN}}$) active power profiles, as well as the relative increase in the \ac{SoC} of the \ac{BESS}, at the \acs{SINTEF} facility. Additionally, it shows the reactive power profiles of \acs{SINTEF} ($Q_\mathrm{el_{SIN}}$) and \acs{RSE} ($Q_\mathrm{el_{RSE}}$), along with the reference frequencies ($f_\mathrm{SIN}^\mathrm{ref}$ and $f_\mathrm{RSE}^\mathrm{ref}$).

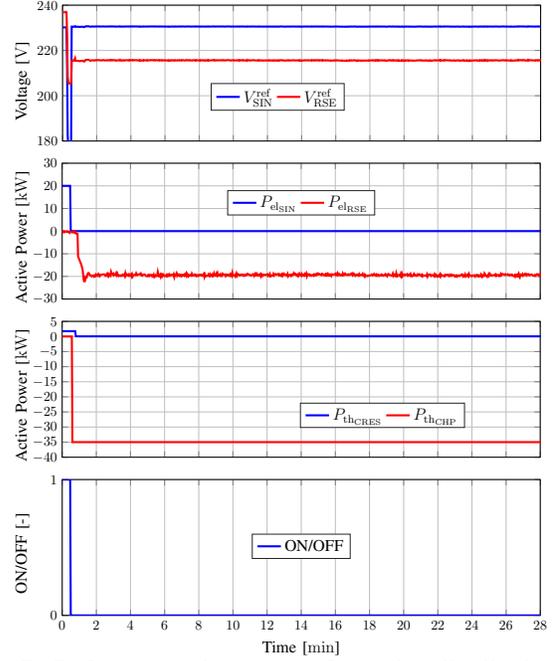
\begin{figure}[tb]
    \centering
    \input{matlabPlots/figure_6_undervoltage.tex}
    \vspace*{-0.65em}
    \caption{Reference voltage sent from the distribution grid (\acs{TUD}) to the converters (\acs{SINTEF} and \acs{RSE}), along with the active power generated by the \ac{BESS} and \ac{CHP}, and the enable signals of the \ac{EHP}, in the undervoltage scenario.}
    \label{fig:underVoltage_1}
\end{figure}

Figure~\ref{fig:underVoltage_1} presents the outcomes of the undervoltage scenario, where the voltage at the coupling points of \acs{SINTEF} and \acs{RSE} suddenly drops below the reference value ($\SI{240}{\volt}$) due to low or zero \ac{PV} generation or high consumption in the grid (see Figure~\ref{fig:CIGRENetwork}). The procedure for this scenario parallels that of the overvoltage scenario, employing a negative threshold value of $\SI{-5}{\%}$ for activation and $\SI{0}{\%}$ for deactivation of units. Initially, the \acs{CRES} \ac{EHP} operates to ensure its deactivation during the experiment, as shown in the lower right corner of Figure~\ref{fig:underVoltage_1}. Hence, the scenario necessitates additional generation by the \ac{CHP}, requiring the utilization of both the \ac{CHP} of \acs{RSE} and the thermal network (\acs{DTU}). 

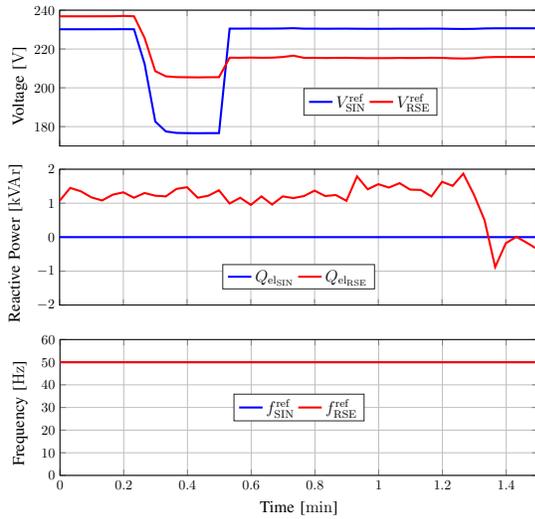
\begin{figure}[tb]
    \centering
    \input{matlabPlots/figure_7_undervoltage.tex}
    \vspace*{-0.65em}
    \caption{Reference voltage, reactive power, and reference frequencies at the \acs{SINTEF} and \acs{RSE} facilities, with a zoomed-in view showing the evolution of the reference voltage in the undervoltage scenario.}
    \label{fig:underVoltage_3}
\end{figure}

\begin{figure}[tb]
    \centering
    \input{matlabPlots/figure_8_undervoltage.tex}
    \vspace*{-0.65em}
    \caption{Power and \acs{SoC} of the \acs{BESS} at \acs{SINTEF}, alongside the temperature of the heat source buffer tank, and the power of heat sent to the buffer, in the undervoltage scenario.}
    \label{fig:underVoltage_2}
\end{figure}
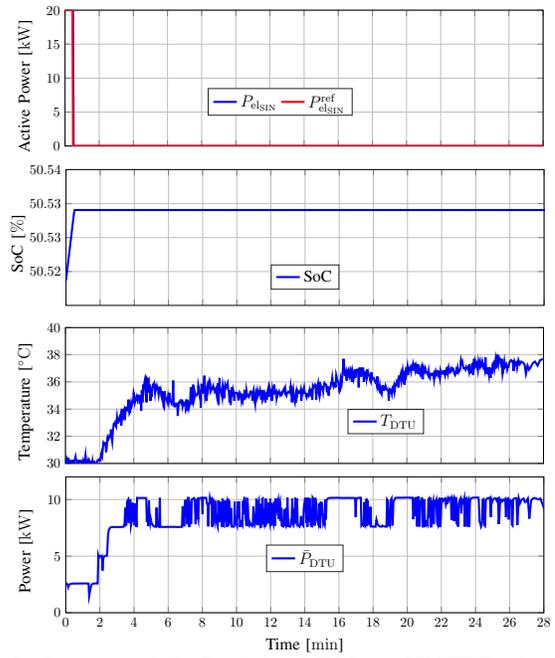

To restore the voltage to the reference value (or close to it), \acs{RSE} generated active power ($P_\mathrm{el_{RSE}}$) by disabling the \ac{EHP} ($P_\mathrm{th_{CRES}}$ goes to zero) and activating the \ac{CHP}, while the \ac{BESS} stopped charging ($P_\mathrm{el_{SIN}}$ goes to zero). This recovery process, depicted in Figure~\ref{fig:underVoltage_3}, occurs within minutes due to the rapid dynamics involved. Notably, the increase in voltage is not directly correlated with active power but instead requires reactive power ($Q_\mathrm{el_{SIN}}$ and $Q_\mathrm{el_{RSE}}$), as evident in the plots (see Figure~\ref{fig:underVoltage_3}). The system utilizes thermal energy to provide services, with insufficient generation of reactive power resulting in the need to generate active power. This supports the fault ride-through capability, enabling the \acp{RI} to remain connected despite voltage fluctuations and provide voltage support and thermal heating services. Figure~\ref{fig:underVoltage_2} demonstrates a decrease in load due to the \ac{BESS} stopping its charging, illustrating the impact of the electrical grid on the thermal network. The battery was maintained at $\SI{50}{\%}$ capacity, as observed in the lower left corner of Figure~\ref{fig:underVoltage_2}. 

The control strategy was simplified, primarily focusing on adjusting active power without considering other parameters like temperatures. If any anomalies arise during the experiments, involved partners could locally override setpoints and deactivate equipment. While both experiments could be combined into one, conducting them separately was more time-efficient, considering the real-time nature of the tests.





\begin{figure}
    \centering
    \includegraphics[width=0.95\columnwidth]{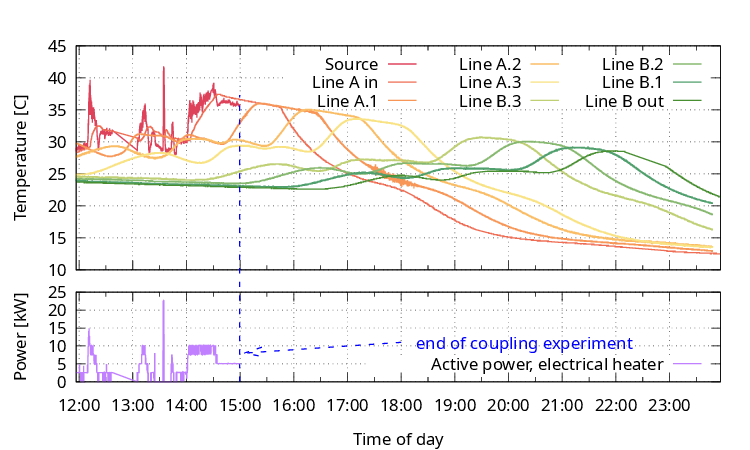}
    \vspace*{-0.65em}
    \caption{Dynamic response of the thermal network along with temperature measurements. 
    }
    \label{fig:thermal_response_all}
\end{figure}

It is important to note that one of the challenges of multi-domain experiments is the difference in the time scales at which relevant phenomena occur. Thermal systems react orders of magnitude more slowly than electrical systems. This is illustrated by the thermal response observed during the experiment, as depicted in Figure~\ref{fig:thermal_response_all}. The bottom plot shows the output of the controllable heat source over a three-hour experiment. The heat source tracks the remote \ac{CHP} plant's output with an offset and scaling factor to match their controllable ranges, limited by a resolution of \SI{2.5}{\kilo\watt} due to the load steps.
The dashed vertical line indicates the end of the coupling experiment, after which no further data is exchanged, and the heat source is turned off. Post-coupling, all dynamic processes occur solely within the thermal system. The upper plot shows dynamic processes continuing for about twelve hours due to the small pump used, which caused low mass flow rates. However, even with a larger pump, the thermal response would still be slower.

Figure~\ref{fig:thermal_response_all} only shows the forward line's response; similar dynamics occur in the return line. Towards the end, cold return water from the heat consumer lowers the temperature at the network start (``Line A in''), as the heat source has been off for hours and the accumulator tank's energy is depleted.

\section{Conclusions}
\label{sec:conclusions}

This paper demonstrated the feasibility and effectiveness of performing~\acl{HIL} and~\acl{GDS} to assess the robustness and responsiveness of multi-energy systems. Through the deployment of the JaNDER middleware and~\ac{uAPI}, secure and efficient communication was maintained across distributed~\acp{RI}, facilitating real-time data exchange and operational control. The experimental demonstration provided insights into the potential to improve grid stability and provide flexibility by integrating power-to-heat services, addressing the critical needs of modern energy systems. 
Future research will focus on refining these technologies and exploring their applications in other multi-energy scenarios to further validate their applicability and impact on grid management practices.


\bibliographystyle{IEEEtran}
\bibliography{main}

\end{document}

%% file: matlabPlots/figure_4_overvoltage.tex
\hspace{-0.735cm}
\begin{subfigure}{0.80\columnwidth}
    \centering
    \scalebox{0.52}{
    \begin{tikzpicture}
    \begin{axis}[%
    width=4.8119in,%
    height=1.3683in,%
    at={(0.758in,0.481in)},%
    scale only axis,%
    xmin=0,%
    xmax=16,%
    ymax=255,%
    ymin=240,%
    xmajorgrids,%
    ymajorgrids,%
    ylabel style={yshift=-0.215cm, xshift=-0.20cm}, 
    xticklabel=\empty,
    ylabel={\large Voltage [\si{\volt}]},%
    axis background/.style={fill=white},%
    legend style={at={(0.45,0.425)},anchor=north,legend cell 
		align=left,draw=none,legend columns=-1,align=left,draw=white!15!black}
    ]
    \addplot[color=blue, solid, line width=1.5pt] 
        file{matlabPlots/overvoltage_attempt1/data_voltage_ref_sin_out_vec_downsample_1.txt};%
    \addplot[color=red, solid, line width=1.5pt] 
        file{matlabPlots/overvoltage_attempt1/data_voltage_ref_rse_out_vec_downsample_1.txt};%
    \legend{\large{$V^\mathrm{ref}_\mathrm{SIN}$}, \large{$V^\mathrm{ref}_\mathrm{RSE}$}};%
    \end{axis}
    \end{tikzpicture}
    }
\end{subfigure}
\\
\vspace{0.05cm}
\hspace{-0.735cm}
\begin{subfigure}{0.80\columnwidth}
    \centering
    \scalebox{0.52}{
    \begin{tikzpicture}
    \begin{axis}[%
    width=4.8119in,%
    height=1.3683in,%
    at={(0.758in,0.481in)},%
    scale only axis,%
    xmin=0,%
    xmax=16,%
    ymax=40,%
    ymin=-40,%
    xmajorgrids,%
    ymajorgrids,%
    ylabel style={yshift=-0.215cm, xshift=-0.15cm}, 
    xticklabel=\empty,
    ylabel={\large Active Power [\si{\kilo\watt}]},%
    axis background/.style={fill=white},%
    legend style={at={(0.50,0.30)}, anchor=north, legend cell align=left, draw=none, legend columns=-1, align=left, draw=white!15!black}
    ]
    \addplot [color=blue, solid, line width=1.5pt] 
        file{matlabPlots/overvoltage_attempt1/data_active_power_el_sin_out_vec_downsample_1.txt};%
    \addplot [color=red, solid, line width=1.5pt] 
        file{matlabPlots/overvoltage_attempt1/data_active_power_el_rse_out_vec_downsample_1.txt};%
    \legend{\large{$P_\mathrm{el_{SIN}}$}, \large{$P_\mathrm{el_{CHP}}$}};%
    \end{axis}
    \end{tikzpicture}
    }
\end{subfigure}
\\
\vspace{0.05cm}
\hspace{-0.735cm}
\begin{subfigure}{0.80\columnwidth}
    \centering
    \scalebox{0.52}{
    \begin{tikzpicture}
    \begin{axis}[%
    width=4.8119in,%
    height=1.3683in,%
    at={(0.758in,0.481in)},%
    scale only axis,%
    xmin=0,%
    xmax=16,%
    ymax=12,%
    ymin=0,%
    xmajorgrids,%
    ymajorgrids,%
    ylabel style={yshift=-0.215cm, xshift=-0.15cm}, 
    xticklabel=\empty,
    ylabel={\large Active Power [\si{\kilo\watt}]},%
    axis background/.style={fill=white},%
    legend style={at={(0.67,0.60)}, anchor=north, legend cell align=left, draw=none, legend columns=-1, align=left, draw=white!15!black}
    ]
    \addplot [color=blue, solid, line width=1.5pt] 
        file{matlabPlots/overvoltage_attempt1/data_cres_heat_pump_m_ea_kiloWatts_vec_downsample_1.txt};%
    \addplot [color=red, solid, line width=1.5pt] 
        file{matlabPlots/overvoltage_attempt1/data_active_power_chp_rse_vec_downsample_1.txt};%
    \legend{\large{$P_\mathrm{th_{CRES}}$}, \large{$P_\mathrm{el_{CHP}}$}};%
    \end{axis}
    \end{tikzpicture}
    }
\end{subfigure}
\\
\vspace{0.05cm}
\hspace{-0.48cm}
\begin{subfigure}{0.80\columnwidth}
    \centering
    \scalebox{0.52}{
    \begin{tikzpicture}
    \begin{axis}[%
    width=4.8119in,%
    height=1.3683in,%
    at={(0.758in,0.481in)},%
    scale only axis,%
    xmin=0,%
    xmax=16,%
    ymax=1,%
    ymin=0,%
    xmajorgrids,%
    ymajorgrids,%
    ylabel style={yshift=-0.215cm, xshift=-0.15cm}, 
    xlabel={\large Time [\si{\minute}]},%
    ylabel={\large ON/OFF [-]},%
    ytick={0,1},%
    axis background/.style={fill=white},%
    legend style={at={(0.50,0.60)}, anchor=north, legend cell align=left, draw=none, legend columns=-1, align=left, draw=white!15!black}
    ]
    \addplot [color=blue, solid, line width=1.5pt] 
        file{matlabPlots/overvoltage_attempt1/data_rse_chp_m_ea_watt_vec_downsample_1.txt};%
    \legend{\large{$\mathrm{ON/OFF}$}};%
    \end{axis}
    \end{tikzpicture}
    }
\end{subfigure}

%% file: matlabPlots/figure_5_overvoltage.tex
\hspace{-0.725cm}
\begin{subfigure}{0.80\columnwidth}
    \centering
    \scalebox{0.52}{
    \begin{tikzpicture}
    \begin{axis}[%
    width=4.8119in,%
    height=1.3683in,%
    at={(0.758in,0.481in)},%
    scale only axis,%
    xmin=0,%
    xmax=16,%
    ymax=40,%
    ymin=-40,%
    xmajorgrids,%
    ymajorgrids,%
    ylabel style={yshift=-0.215cm, xshift=-0.20cm}, 
    xticklabel=\empty,
    ylabel={\large Active Power [\si{\kilo\watt}]},%
    axis background/.style={fill=white},%
    legend style={at={(0.45,0.425)},anchor=north,legend cell 
		align=left,draw=none,legend columns=-1,align=left,draw=white!15!black}
    ]
    \addplot[color=blue, solid, line width=1.5pt] 
        file{matlabPlots/overvoltage_attempt1/data_active_power_el_sin_out_vec_downsample_1.txt};%
    \addplot[color=red, solid, line width=1.5pt] 
        file{matlabPlots/overvoltage_attempt1/data_active_power_ref_el_sin_out_vec_downsample_1.txt};%
    \legend{\large{$P_\mathrm{el_{SIN}}$}, \large{$P^\mathrm{ref}_\mathrm{el_{SIN}}$}};%
    \end{axis}
    \end{tikzpicture}
    }
\end{subfigure}
\\
\vspace{0.05cm}
\hspace{-0.735cm}
\begin{subfigure}{0.80\columnwidth}
    \centering
    \scalebox{0.52}{
    \begin{tikzpicture}
    \begin{axis}[%
    width=4.8119in,%
    height=1.3683in,%
    at={(0.758in,0.481in)},%
    scale only axis,%
    xmin=0,%
    xmax=16,%
    ymax=50.4,%
    ymin=49.9,%
    xmajorgrids,%
    ymajorgrids,%
    ylabel style={yshift=-0.215cm, xshift=-0.15cm}, 
    xticklabel=\empty,
    ylabel={\large SoC [\si{\%}]},%
    axis background/.style={fill=white},%
    legend style={at={(0.50,0.30)}, anchor=north, legend cell align=left, draw=none, legend columns=-1, align=left, draw=white!15!black}
    ]
    \addplot [color=blue, solid, line width=1.5pt] 
        file{matlabPlots/overvoltage_attempt1/data_soc_sin_vec_downsample_1.txt};%
    \legend{\large{$\mathrm{SoC}$}};%
    \end{axis}
    \end{tikzpicture}
    }
\end{subfigure}
\\
\vspace{0.05cm}
\hspace{-0.735cm}
\begin{subfigure}{0.80\columnwidth}
    \centering
    \scalebox{0.52}{
    \begin{tikzpicture}
    \begin{axis}[%
    width=4.8119in,%
    height=1.3683in,%
    at={(0.758in,0.481in)},%
    scale only axis,%
    xmin=0,%
    xmax=16,%
    ymax=0.55,%
    ymin=-0.55,%
    xmajorgrids,%
    ymajorgrids,%
    ylabel style={yshift=-0.215cm, xshift=-0.15cm}, 
    xticklabel=\empty,
    ylabel={\large Reactive Power [\si{\kilo\volt\ampere r}]},%
    axis background/.style={fill=white},%
    legend style={at={(0.67,0.40)}, anchor=north, legend cell align=left, draw=none, legend columns=-1, align=left, draw=white!15!black}
    ]
    \addplot [color=blue, solid, line width=1.5pt] 
        file{matlabPlots/overvoltage_attempt1/data_reactive_power_el_sin_out_vec_downsample_1.txt};%
    \addplot [color=red, solid, line width=1.5pt] 
        file{matlabPlots/overvoltage_attempt1/data_reactive_power_el_rse_out_vec_downsample_1.txt};%
    \legend{\large{$Q_\mathrm{el_{SIN}}$}, \large{$Q_\mathrm{el_{RSE}}$}};%
    \end{axis}
    \end{tikzpicture}
    }
\end{subfigure}
\\
\vspace{0.05cm}
\hspace{-0.48cm}
\begin{subfigure}{0.80\columnwidth}
    \centering
    \scalebox{0.52}{
    \begin{tikzpicture}
    \begin{axis}[%
    width=4.8119in,%
    height=1.3683in,%
    at={(0.758in,0.481in)},%
    scale only axis,%
    xmin=0,%
    xmax=16,%
    ymax=60,%
    ymin=0,%
    xmajorgrids,%
    ymajorgrids,%
    ylabel style={yshift=-0.215cm, xshift=-0.15cm}, 
    xlabel={\large Time [\si{\minute}]},%
    ylabel={\large Frequency [\si{\hertz}]},%
    ytick={0,10,20,30,40,50,60},%
    axis background/.style={fill=white},%
    legend style={at={(0.50,0.60)}, anchor=north, legend cell align=left, draw=none, legend columns=-1, align=left, draw=white!15!black}
    ]
    \addplot [color=blue, solid, line width=1.5pt] 
        file{matlabPlots/overvoltage_attempt1/data_frequency_ref_sin_out_vec_downsample_1.txt};%
    \addplot [color=red, solid, line width=1.5pt] 
        file{matlabPlots/overvoltage_attempt1/data_frequency_ref_rse_out_vec_downsample_1.txt};%
    \legend{\large{$f^\mathrm{ref}_\mathrm{SIN}$}, \large{$f^\mathrm{ref}_\mathrm{RSE}$}};%
    \end{axis}
    \end{tikzpicture}
    }
\end{subfigure}

%% file: matlabPlots/figure_6_undervoltage.tex
\hspace{-0.735cm}
\begin{subfigure}{0.80\columnwidth}
    \centering
    \scalebox{0.52}{
    \begin{tikzpicture}
    \begin{axis}[%
    width=4.8119in,%
    height=1.3683in,%
    at={(0.758in,0.481in)},%
    scale only axis,%
    xmin=0,%
    xmax=28,%
    ymax=240,%
    ymin=180,%
    xmajorgrids,%
    ymajorgrids,%
    ylabel style={yshift=-0.215cm, xshift=-0.20cm}, 
    xticklabel=\empty,
    ylabel={\large Voltage [\si{\volt}]},%
    axis background/.style={fill=white},%
    legend style={at={(0.45,0.425)},anchor=north,legend cell 
		align=left,draw=none,legend columns=-1,align=left,draw=white!15!black}
    ]
    \addplot[color=blue, solid, line width=1.5pt] 
        file{matlabPlots/undervoltage/data_voltage_ref_sin_out_vec_downsample_1.txt};%
    \addplot[color=red, solid, line width=1.5pt] 
        file{matlabPlots/undervoltage/data_voltage_ref_rse_out_vec_downsample_1.txt};%
    \legend{\large{$V^\mathrm{ref}_\mathrm{SIN}$}, \large{$V^\mathrm{ref}_\mathrm{RSE}$}};%
    \end{axis}
    \end{tikzpicture}
    }
\end{subfigure}
\\
\vspace{0.05cm}
\hspace{-0.735cm}
\begin{subfigure}{0.80\columnwidth}
    \centering
    \scalebox{0.52}{
    \begin{tikzpicture}
    \begin{axis}[%
    width=4.8119in,%
    height=1.3683in,%
    at={(0.758in,0.481in)},%
    scale only axis,%
    xmin=0,%
    xmax=28,%
    ymax=30,%
    ymin=-30,%
    xmajorgrids,%
    ymajorgrids,%
    ylabel style={yshift=-0.215cm, xshift=-0.15cm}, 
    xticklabel=\empty,
    ylabel={\large Active Power [\si{\kilo\watt}]},%
    ytick={30,20,10,0,-10,-20,-30},%
    axis background/.style={fill=white},%
    legend style={at={(0.50,0.80)}, anchor=north, legend cell align=left, draw=none, legend columns=-1, align=left, draw=white!15!black}
    ]
    \addplot [color=blue, solid, line width=1.5pt] 
        file{matlabPlots/undervoltage/data_active_power_el_sin_out_vec_downsample_1.txt};%
    \addplot [color=red, solid, line width=1.5pt] 
        file{matlabPlots/undervoltage/data_active_power_el_rse_out_vec_downsample_1.txt};%
    \legend{\large{$P_\mathrm{el_{SIN}}$}, \large{$P_\mathrm{el_{RSE}}$}};%
    \end{axis}
    \end{tikzpicture}
    }
\end{subfigure}
\\
\vspace{0.05cm}
\hspace{-0.735cm}
\begin{subfigure}{0.80\columnwidth}
    \centering
    \scalebox{0.52}{
    \begin{tikzpicture}
    \begin{axis}[%
    width=4.8119in,%
    height=1.3683in,%
    at={(0.758in,0.481in)},%
    scale only axis,%
    xmin=0,%
    xmax=28,%
    ymax=5,%
    ymin=-40,%
    xmajorgrids,%
    ymajorgrids,%
    ylabel style={yshift=-0.215cm, xshift=-0.15cm}, 
    xticklabel=\empty,
    ylabel={\large Active Power [\si{\kilo\watt}]},%
    ytick={5,0,...,-40},%
    axis background/.style={fill=white},%
    legend style={at={(0.67,0.40)}, anchor=north, legend cell align=left, draw=none, legend columns=-1, align=left, draw=white!15!black}
    ]
    \addplot [color=blue, solid, line width=1.5pt] 
        file{matlabPlots/undervoltage/data_cres_heat_pump_m_ea_kiloWatts_vec_downsample_1.txt};%
    \addplot [color=red, solid, line width=1.5pt] 
        file{matlabPlots/undervoltage/data_active_power_chp_rse_vec_downsample_1.txt};%
    \legend{\large{$P_\mathrm{th_{CRES}}$}, \large{$P_\mathrm{th_{CHP}}$}};%
    \end{axis}
    \end{tikzpicture}
    }
\end{subfigure}
\\
\vspace{0.05cm}
\hspace{-0.485cm}
\begin{subfigure}{0.80\columnwidth}
    \centering
    \scalebox{0.52}{
    \begin{tikzpicture}
    \begin{axis}[%
    width=4.8119in,%
    height=1.3683in,%
    at={(0.758in,0.481in)},%
    scale only axis,%
    xmin=0,%
    xmax=28,%
    ymax=1,%
    ymin=0,%
    xmajorgrids,%
    ymajorgrids,%
    ylabel style={yshift=-0.215cm, xshift=-0.15cm}, 
    xlabel={\large Time [\si{\minute}]},%
    ylabel={\large ON/OFF [-]},%
    ytick={0,1},%
    axis background/.style={fill=white},%
    legend style={at={(0.50,0.60)}, anchor=north, legend cell align=left, draw=none, legend columns=-1, align=left, draw=white!15!black}
    ]
    \addplot [color=blue, solid, line width=1.5pt] 
        file{matlabPlots/undervoltage/data_rse_chp_m_ea_watt_vec_downsample_1.txt};%
    \legend{\large{ON/OFF}};%
    \end{axis}
    \end{tikzpicture}
    }
\end{subfigure}

%% file: matlabPlots/figure_7_undervoltage.tex
\hspace{-0.525cm}
\begin{subfigure}{0.80\columnwidth}
    \centering
    \scalebox{0.52}{
    \begin{tikzpicture}
    \begin{axis}[%
    width=4.8119in,%
    height=1.3683in,%
    at={(0.758in,0.481in)},%
    scale only axis,%
    xmin=0,%
    xmax=1.5,%
    ymax=240,%
    ymin=170,%
    xmajorgrids,%
    ymajorgrids,%
    ylabel style={yshift=-0.215cm, xshift=-0.20cm}, 
    xticklabel=\empty,
    ylabel={\large Voltage [\si{\volt}]},%
    axis background/.style={fill=white},%
    legend style={at={(0.65,0.425)},anchor=north,legend cell 
		align=left,draw=none,legend columns=-1,align=left,draw=white!15!black}
    ]
    \addplot[color=blue, solid, line width=1.5pt] 
        file{matlabPlots/undervoltage/data_voltage_ref_sin_out_vec_downsample_1.txt};%
    \addplot[color=red, solid, line width=1.5pt] 
        file{matlabPlots/undervoltage/data_voltage_ref_rse_out_vec_downsample_1.txt};%
    \legend{\large{$V^\mathrm{ref}_\mathrm{SIN}$}, \large{$V^\mathrm{ref}_\mathrm{RSE}$}};%
    \end{axis}
    \end{tikzpicture}
    }
\end{subfigure}
\\
\vspace{0.05cm}
\hspace{-0.635cm}
\begin{subfigure}{0.80\columnwidth}
    \centering
    \scalebox{0.52}{
    \begin{tikzpicture}
    \begin{axis}[%
    width=4.8119in,%
    height=1.3683in,%
    at={(0.758in,0.481in)},%
    scale only axis,%
    xmin=0,%
    xmax=1.5,%
    ymax=2,%
    ymin=-2,%
    xmajorgrids,%
    ymajorgrids,%
    ylabel style={yshift=-0.1cm, xshift=-0.15cm}, 
    xticklabel=\empty,
    ylabel={\large Reactive Power [\si{\kilo\volt\ampere r}]},%
    axis background/.style={fill=white},%
    legend style={at={(0.50,0.30)}, anchor=north, legend cell align=left, draw=none, legend columns=-1, align=left, draw=white!15!black}
    ]
    \addplot [color=blue, solid, line width=1.5pt] 
        file{matlabPlots/undervoltage/data_reactive_power_el_sin_out_vec_downsample_1.txt};%
    \addplot [color=red, solid, line width=1.5pt] 
        file{matlabPlots/undervoltage/data_reactive_power_el_rse_out_vec_downsample_1.txt};%
    \legend{\large{$Q_\mathrm{el_{SIN}}$}, \large{$Q_\mathrm{el_{RSE}}$}};%
    \end{axis}
    \end{tikzpicture}
    }
\end{subfigure}
\\
\vspace{0.05cm}
\hspace{-0.27cm}
\begin{subfigure}{0.80\columnwidth}
    \centering
    \scalebox{0.52}{
    \begin{tikzpicture}
    \begin{axis}[%
    width=4.8119in,%
    height=1.3683in,%
    at={(0.758in,0.481in)},%
    scale only axis,%
    xmin=0,%
    xmax=1.5,%
    ymax=60,%
    ymin=0,%
    xmajorgrids,%
    ymajorgrids,%
    ylabel style={yshift=-0.215cm, xshift=-0.15cm}, 
    xlabel={\large Time [\si{\minute}]},%
    ylabel={\large Frequency [\si{\hertz}]},%
    ytick={0,10,20,30,40,50,60},%
    axis background/.style={fill=white},%
    legend style={at={(0.50,0.60)}, anchor=north, legend cell align=left, draw=none, legend columns=-1, align=left, draw=white!15!black}
    ]
    \addplot [color=blue, solid, line width=1.5pt] 
        file{matlabPlots/undervoltage/data_frequency_ref_sin_out_vec_downsample_1.txt};%
    \addplot [color=red, solid, line width=1.5pt] 
        file{matlabPlots/undervoltage/data_frequency_ref_rse_out_vec_downsample_1.txt};%
    \legend{\large{$f^\mathrm{ref}_\mathrm{SIN}$}, \large{$f^\mathrm{ref}_\mathrm{RSE}$}};%
    \end{axis}
    \end{tikzpicture}
    }
\end{subfigure}

%% file: matlabPlots/figure_8_undervoltage.tex
\hspace{-0.525cm}
\begin{subfigure}{0.80\columnwidth}
    \centering
    \scalebox{0.52}{
    \begin{tikzpicture}
    \begin{axis}[%
    width=4.8119in,%
    height=1.3683in,%
    at={(0.758in,0.481in)},%
    scale only axis,%
    xmin=0,%
    xmax=28,%
    ymax=20,%
    ymin=0,%
    xmajorgrids,%
    ymajorgrids,%
    ylabel style={yshift=-0.215cm, xshift=-0.20cm}, 
    xticklabel=\empty,
    ylabel={\large Active Power [\si{\kilo\watt}]},%
    axis background/.style={fill=white},%
    legend style={at={(0.45,0.425)},anchor=north,legend cell 
		align=left,draw=none,legend columns=-1,align=left,draw=white!15!black}
    ]
    \addplot[color=blue, solid, line width=1.5pt] 
        file{matlabPlots/undervoltage/data_active_power_el_sin_out_vec_downsample_1.txt};%
    \addplot[color=red, solid, line width=1.5pt] 
        file{matlabPlots/undervoltage/data_active_power_ref_el_sin_out_vec_downsample_1.txt};%
    \legend{\large{$P_\mathrm{el_{SIN}}$}, \large{$P^\mathrm{ref}_\mathrm{el_{SIN}}$}};%
    \end{axis}
    \end{tikzpicture}
    }
\end{subfigure}
\\
\vspace{0.05cm}
\hspace{-0.735cm}
\begin{subfigure}{0.80\columnwidth}
    \centering
    \scalebox{0.52}{
    \begin{tikzpicture}
    \begin{axis}[%
    width=4.8119in,%
    height=1.3683in,%
    at={(0.758in,0.481in)},%
    scale only axis,%
    xmin=0,%
    xmax=28,%
    ymax=50.54,%
    ymin=50.52,%
    xmajorgrids,%
    ymajorgrids,%
    ylabel style={yshift=-0cm, xshift=-0.15cm}, 
    xticklabel=\empty,
    ylabel={\large SoC [\si{\%}]},%
    axis background/.style={fill=white},%
    legend style={at={(0.50,0.30)}, anchor=north, legend cell align=left, draw=none, legend columns=-1, align=left, draw=white!15!black}
    ]
    \addplot [color=blue, solid, line width=1.5pt] 
        file{matlabPlots/undervoltage/data_soc_sin_vec_downsample_1.txt};%
    \legend{\large{SoC}};%
    \end{axis}
    \end{tikzpicture}
    }
\end{subfigure}
\\
\vspace{0.05cm}
\hspace{-0.525cm}
\begin{subfigure}{0.80\columnwidth}
    \centering
    \scalebox{0.52}{
    \begin{tikzpicture}
    \begin{axis}[%
    width=4.8119in,%
    height=1.3683in,%
    at={(0.758in,0.481in)},%
    scale only axis,%
    xmin=0,%
    xmax=28,%
    ymax=40,%
    ymin=30,%
    xmajorgrids,%
    ymajorgrids,%
    ylabel style={yshift=-0.215cm, xshift=-0.15cm}, 
    xticklabel=\empty,
    ylabel={\large Temperature [\si{\degreeCelsius}]},%
    axis background/.style={fill=white},%
    legend style={at={(0.67,0.40)}, anchor=north, legend cell align=left, draw=none, legend columns=-1, align=left, draw=white!15!black}
    ]
    \addplot [color=blue, solid, line width=1.5pt] 
        file{matlabPlots/undervoltage/data_th_temp_0_dtu_vec_downsample_1.txt};%
    \legend{\large{$T_\mathrm{DTU}$}};%
    \end{axis}
    \end{tikzpicture}
    }
\end{subfigure}
\\
\vspace{0.05cm}
\hspace{-0.26cm}
\begin{subfigure}{0.80\columnwidth}
    \centering
    \scalebox{0.52}{
    \begin{tikzpicture}
    \begin{axis}[%
    width=4.8119in,%
    height=1.3683in,%
    at={(0.758in,0.481in)},%
    scale only axis,%
    xmin=0,%
    xmax=28,%
    ymax=12,%
    ymin=0,%
    xmajorgrids,%
    ymajorgrids,%
    ylabel style={yshift=-0.215cm, xshift=-0.15cm}, 
    xlabel={\large Time [\si{\minute}]},%
    ylabel={\large Power [\si{\kilo\watt}]},%
    axis background/.style={fill=white},%
    legend style={at={(0.50,0.50)}, anchor=north, legend cell align=left, draw=none, legend columns=-1, align=left, draw=white!15!black}
    ]
    \addplot [color=blue, solid, line width=1.5pt] 
        file{matlabPlots/undervoltage/data_thermal_source_dtu_vec_downsample_1.txt};%
    %
    \legend{\large{$\bar{P}_\mathrm{{DTU}}$}};%
    \end{axis}
    \end{tikzpicture}
    }
\end{subfigure}